\documentclass{elsart3p}

\usepackage{epsfig}

\usepackage{amssymb}

\begin{document}

\begin{frontmatter}

% Title, authors and addresses

% use the thanksref command within \title, \author or \address for footnotes;
% use the corauthref command within \author for corresponding author footnotes;
% use the ead command for the email address,
% and the form \ead[url] for the home page:
% \title{Title\thanksref{label1}}
% \thanks[label1]{}
% \author{Name\corauthref{cor1}\thanksref{label2}}
% \ead{email address}
% \ead[url]{home page}
% \thanks[label2]{}
% \corauth[cor1]{}
% \address{Address\thanksref{label3}}
% \thanks[label3]{}

\title{Measurement of CP asymmetry in Cabibbo suppressed $D^0$ decays}

% use optional labels to link authors explicitly to addresses:
% \author[label1,label2]{}
% \address[label1]{}
% \address[label2]{}

%%% Paper:    CP asymmetry in Cabibbo-suppressed D0 decays
%%% Journal:  Physics Letters B
%%% Contacts: M. Staric (marko.staric@ijs.si)
%%% Use \input{author} to insert this material into your latex file.
\collab{Belle Collaboration}
  \author[JSI]{M.~Stari\v c}, % Ljubljana
  \author[KEK]{I.~Adachi}, % KEK
  \author[Tokyo]{H.~Aihara}, % Tokyo
  \author[BINP]{K.~Arinstein}, % BINP
  \author[Lausanne,ITEP]{T.~Aushev}, % ITEP
  \author[Sydney]{A.~M.~Bakich}, % Sydney
  \author[Melbourne]{E.~Barberio}, % Melbourne
  \author[Lausanne]{A.~Bay}, % Lausanne
  \author[BINP]{I.~Bedny}, % BINP
  \author[Protvino]{K.~Belous}, % Protvino
% ...
% BINP

\address[BINP]{Budker Institute of Nuclear Physics, Novosibirsk, Russia}
% ...
\address[KEK]{High Energy Accelerator Research Organization (KEK), Tsukuba, Japan}
% ...
\address[Protvino]{Institute for High Energy Physics, Protvino, Russia}
% ...
\address[ITEP]{Institute for Theoretical and Experimental Physics, Moscow, Russia}
\address[JSI]{J. Stefan Institute, Ljubljana, Slovenia}
% ...
\address[Lausanne]{\'Ecole Polytechnique F\'ed\'erale de Lausanne, EPFL, Lausanne, Switzerland}
% ...
\address[Melbourne]{University of Melbourne, Victoria, Australia}
% ...
\address[Sydney]{University of Sydney, Sydney, NSW, Australia}
% ...
\address[Tokyo]{Department of Physics, University of Tokyo, Tokyo, Japan}
% ...

\begin{abstract}
  We measure the $CP$-violating asymmetries in decays to the
  $D^0 \to K^+K^-$ and $D^0 \to \pi^+\pi^-$ $CP$ eigenstates
  using 540~fb$^{-1}$ of data collected with the Belle detector at or near 
  the $\Upsilon(4S)$ resonance. Cabibbo-favored $D^0 \to K^-\pi^+$ decays are
  used to correct for systematic detector effects. The results,
  $A_{CP}^{KK} = (-0.43 \pm 0.30 \pm 0.11)\%$ and
  $A_{CP}^{\pi\pi} = (+0.43 \pm 0.52 \pm 0.12)\%$,
  are consistent with no $CP$ violation.
\end{abstract}

\end{frontmatter}

\section{Introduction}
\label{sec1}

Decays of neutral $D$ mesons are a promising area in which to search
for physics beyond the Standard Model (SM). Recently, 
evidence for mixing in this system has been obtained~\cite{belle_mix,babar_mix,cdf_mix}. 
However, whether the effect observed 
is due to the Cabibbo-Kobayashi-Maskawa (CKM) theory or due to new physics (NP)
has yet to be determined and will require further measurements to
resolve. One possible measurement sensitive to NP is that of
a $CP$ asymmetry in $D^0$ decays to Cabibbo-suppressed (CS) final
states~\cite{cpv_th_scs}. Within the SM such an asymmetry is predicted to be very
small ($\lesssim 0.1$\%), but within NP scenarios it can be substantial
($\gtrsim 1$\%)~\cite{cpv_th_scs,cpv_th}.  

In this Letter we present a high statistics search for a $CP$ 
asymmetry in the CS modes $D^0\to K^+K^-$ and $D^0\to \pi^+\pi^-$. 
These final states are accessible to both $D^0$ and $\bar{D}^0$ mesons.
The time-integrated $CP$ asymmetry for decays into a $CP$
eigenstate $f$ is defined as
\begin{equation}
  A_{CP}^f=\frac{\Gamma(D^0\to f)-\Gamma(\bar{D}^0\to f)}{\Gamma(D^0\to
    f)+\Gamma(\bar{D}^0\to f)}=a_d^f+a_{\rm ind}~.
  \label{eq1}
\end{equation}
This quantity receives contributions from both direct ($a_d^f$) and 
indirect ($a_{\rm ind}$) $CP$ violation (CPV)~\cite{cpv_th_scs}. 
While the direct contribution is in general 
distinct for different final states, the indirect contribution
is the same. The indirect CPV contribution is constrained by our recent measurement 
of the lifetime difference using $D^0(\bar{D}^0)\to K^+K^-,\pi^+\pi^-$ 
decays~\cite{belle_mix}: $A_\Gamma\equiv -a_{\rm ind}=(0.01\pm 0.30\pm 0.15)\%$.  
$CP$ asymmetries in CS decays have been searched for previously using
$D^0\to K^+K^-,\pi^+\pi^-$~\cite{babar_acp} and 
$D^0\to \pi^+\pi^-\pi^0, K^+K^-\pi^0$~\cite{acp_pipipi}.

\section{Method}
\label{sec2}

The flavour of neutral $D$ mesons at production is tagged by
reconstructing $D^{\ast +}\to D^0\pi_s^+$ decays\footnote{Charge
  conjugated processes are implied throughout the paper, unless
  explicitly noted otherwise.} in which the charge of the low momentum
pion, $\pi_s$, determines the flavour of the $D^0$ meson. 
The measured asymmetry, 
$A_{\rm rec}^f=[N(D^0\to f)-N(\bar{D}^0\to f)]/[N(D^0\to f)+N(\bar{D}^0\to f)]$, 
with $f=K^+K^-,\pi^+\pi^-$ and $N$ denoting the number of reconstructed decays,
can be written as a sum of several (assumed small) contributions:
\begin{equation}
  A_{\rm rec}^f=A_{FB}+A_{CP}^f+A_\epsilon^\pi~.
  \label{eq2}
\end{equation}
In addition to the intrinsic asymmetry $A_{CP}^f$ there is a contribution due
to an asymmetry in the reconstruction efficiencies of oppositely
charged $\pi_s$ ($A_\epsilon^\pi$). 
Since the final state $f$ is self-conjugate, its
reconstruction efficiency does not affect $A_{\rm rec}^f$.
Furthermore, there is a
forward-backward asymmetry ($A_{FB}$) in the production of $D^{\ast +}$ mesons
in $e^+e^-\to c\bar{c}$ arising from $\gamma-Z^0$ interference and
higher order QED effects~\cite{afb}.  
This term is an odd function of the cosine of the $D^{\ast +}$ production polar
angle in the center-of-mass (CM) system\footnote{Symbols with an asterix 
  in the paper denote quantities
  in the CM frame, while those without asterix denote quantities in the
  laboratory frame.} ($\cos\theta^\ast$). Since our detector acceptance
is not symmetric with respect to $\cos\theta^\ast$,
the measurement is performed in bins of $\cos\theta^\ast$. 
This allows us to correct for acceptance and extract
both $A_{FB}$ and $A_{CP}^f$ as described below.

To reliably determine $A_\epsilon^\pi$ we adopt the 
method of Ref.~\cite{babar_acp} with some appropriate modifications. 
In addition to the $D^0\to h^+h^-$ modes mentioned above, we
also reconstruct two $D^0\to K^-\pi^+$ samples: one consisting of
$D$ mesons with tagged initial flavour, and one consisting
of untagged candidates. The measured asymmetries for these
modes can be written as
\begin{eqnarray}
  \nonumber
  A_{\rm rec}^{\rm tag}~~~&=&A_{FB}+A_{CP}^{K\pi}+A_\epsilon^{K\pi}+A_\epsilon^\pi~,\\
  A_{\rm rec}^{\rm untag}&=&A_{FB}+A_{CP}^{K\pi}+A_\epsilon^{K\pi}~.
  \label{eq3}
\end{eqnarray}
A notable difference with (\ref{eq2}) is that this final state is
not self-conjugate and thus an additional term $A_\epsilon^{K\pi}$
appears as a consequence of a possible asymmetry
in the reconstruction efficiency. We first use the two
measurements in (\ref{eq3}) to determine $A_\epsilon^\pi$;
we then insert the result into (\ref{eq2}) and use the fact
that $A_{FB}$ is antisymmetric with respect to $\cos\theta^\ast$ and $A_{CP}^f$ is
independent of this variable.

Reconstruction efficiencies and their asymmetries  $A_\epsilon^i$,
however, are functions of momenta of particles $i=\pi_s,~ K\pi$ in the
laboratory frame. For a $D^0$ meson with a given momentum
$\vec{p}_{D^0}$, the efficiency of reconstructing the final state
$K^-\pi^+$ is $\epsilon_{K\pi}(\vec{p}_{D^0})=
\int\epsilon_K(\vec{p}_K) \epsilon_\pi(\vec{p}_\pi)
w_{\vec{p}_{D^0}}(\vec{p}_K,\vec{p}_\pi)d\vec{p}_Kd\vec{p}_\pi$, where
$w_{\vec{p}_{D^0}}(\vec{p}_K,\vec{p}_\pi)$ denotes the 6-dimensional
distribution of final state particles. For a given $\vec{p}_{D^0}$, 
this distribution is independent
of whether the $D$ meson candidate is flavour-tagged or not.
Using the same selection criteria for the $K$ and $\pi$
candidates in the tagged and untagged sample imposes equality of the
selection efficiencies $\epsilon_{K(\pi)}(\vec{p}_{K(\pi)})$. Hence
the asymmetry $A_\epsilon^{K\pi}(\vec{p}_{D^0})$ is identical for
tagged and untagged $D$ mesons of a given momentum $\vec{p}_{D^0}$,
as implied by (\ref{eq3}). Since the distribution of $D^0$
mesons is uniform in the azimuthal angle the dimension of the problem can be reduced. 
It is sufficient to obtain
$A_\epsilon^{K\pi}$ as a function of the magnitude and polar angle
of the laboratory momentum, $p_{D^0}$ and $\cos\theta_{D^0}$. 

The slow pion asymmetry $A_\epsilon^\pi$ depends on its momentum
$\vec{p}_{\pi_s}$ and is independent of the $D^0$ final state. Since
the $\pi_s$ azimuthal angle distribution is also found to be uniform,
$A_\epsilon^\pi$ is examined as a function of $(p_{\pi_s},\cos\theta_{\pi_s})$.

\section{Measurement}
\label{sec3}

The measurement is based on 540~fb$^{-1}$ of data recorded
by the Belle detector~\cite{Belle} at the KEKB asymmetric-energy
$e^+e^-$~collider~\cite{KEKB}, running at the CM energy of
the $\Upsilon(4S)$ resonance and 60~MeV below.
The Belle detector is described in detail elsewhere~\cite{Belle}:
it includes in particular a silicon vertex detector (SVD), a
central drift chamber,
an array of aerogel Cherenkov counters,
and time-of-flight scintillation counters.
Two different SVD configurations were used: a 3-layer
configuration for the first 153~fb$^{-1}$ of data, and a 4-layer configuration~\cite{SVD2}
for the remaining data.

We reconstruct $D^{\ast +}\to D^0\pi^+_s$, $D^0\to K^+K^-$, $K^-\pi^+$,
$\pi^+\pi^-$ decay chains, as well as the decay $D^0 \to K^-\pi^+$
without requiring an accompanying $D^{\ast +}$ decay.  
Each final state charged particle is required
to have at least two associated SVD hits in each of the two
measuring coordinates.
To select pion and kaon candidates, we impose
standard particle identification criteria~\cite{PID}.
$D^0$ daughter particles are refitted to a common vertex. 
The $D^0$ production vertex is found by constraining the $D^0$ (and $\pi_s$ for
the tagged decays) to originate from the $e^+e^-$ interaction region.
Confidence levels exceeding $10^{-3}$
are required for both fits.  The $D^{\ast +}$ ($D^0$ for untagged decays) momentum 
must satisfy $p^\ast_D>~2.5$~GeV/$c^2$ in order to reject $D$-mesons
produced in $B$-meson decays and to suppress combinatorial background.

\begin{figure}
  \centerline{\epsfig{file=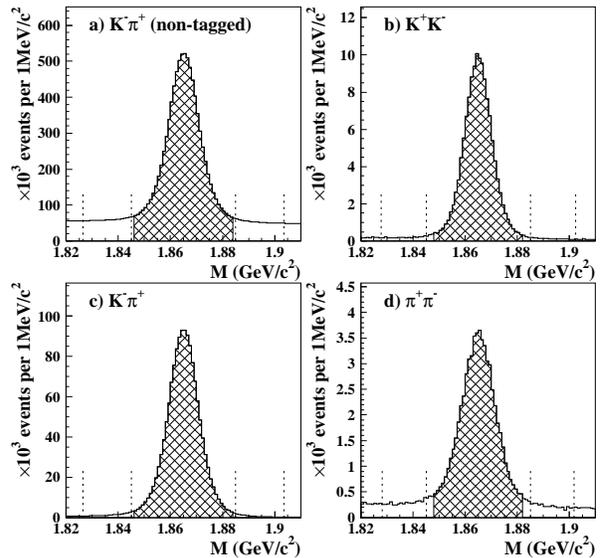,width=8cm}}
  \caption{Invariant mass spectra of selected events.
    For the tagged data samples (b,c,d) events with $|\Delta q|<$1~MeV are selected. 
    The cross-hatched area represents the signal region; the sideband positions
    are indicated by vertical lines.
  }
  \label{mass_spectra.eps}
\end{figure}

We accept candidates with a $D^0$ invariant mass $M$ in
the range $1.81~{\rm GeV/c}^2~ < M < 1.91~{\rm GeV/c}^2$. For final
states with a $\pi_s$, we require that the energy released
in the $D^{*+}$ decay, $q=(M_{D^{*+}}-M-m_\pi)c^2$, be less than 20~MeV.
In this expression, $M_{D^{*+}}$ is the invariant mass of the
$D^0\pi_s^+$ combination and $m_\pi$ is the charged pion mass.
For the small fraction of events with multiple
candidates (0.1\% for the tagged samples, 2.9\%
for the untagged sample), we select only one
candidate: that in which the sum of the production
and decay vertex $\chi^2$'s is smallest. 
We also require\footnote{This cut limits the range of 
  measurement to $|\cos \theta^*|<0.8$} 
$|\cos \theta_{D^0}|<0.9$ to remove 
events in which large slow pion asymmetry corrections and
consequently large systematic uncertainties are expected.
The resulting invariant mass spectra are shown in Fig.~\ref{mass_spectra.eps}. 

We measure the signal yield by performing a mass-sideband subtraction, as this method is
robust and reduces sensitivity to the signal shape.
The possibility of a non-linear background shape 
is considered as a systematic uncertainty.
The sizes of signal windows in $M$ and $q$ are chosen to minimize
the expected statistical error on the $A_{CP}$ measurement. Using the Monte Carlo (MC)
simulation, which has been tuned to reproduce the signal shapes and
the signal-to-background ratios of the real data, 
the optimal signal windows are found to be
$|\Delta M| < 17.3~(18.6,16.8)$~MeV/c$^2$ and $|\Delta q| < 1.00~(1.85,0.90)$~MeV for
the $KK~(K\pi,\pi\pi)$ final states.
The quantities $\Delta M$ and $\Delta q$ measure the difference of the
corresponding observable and the nominal 
$D^0$ mass and the nominal released energy of the $D^{*+}$ decay, respectively.
Sidebands of the same size as signal window
are chosen starting at $\pm$20~MeV/$c^2$ from the $D^0$ nominal mass.
Within the optimal signal window we find $6.3\times 10^6$ untagged $K^-\pi^+$ 
signal events with a purity of 80\%; 
the number of tagged signal events is 
$120\times 10^3~K^+K^-$, $1.3\times 10^6~K^-\pi^+$ and 
$51\times 10^3~\pi^+\pi^-$, with purities of 97\%, 99\% and 91\%, respectively.

We determine first the asymmetry $A_{\rm rec}^{\rm untag}$ of the untagged $K\pi$ sample 
in $20\times 20$ bins of the two-dimensional phase space $(p_{D^0}, \cos \theta_{D^0})$
by
\begin{equation}
  A_{{\rm rec},ij}^{\rm untag} = \frac{N_{ij}-\overline{N}_{ij}}{N_{ij}+\overline{N}_{ij}}~,
\end{equation}
where $N_{ij}$ and $\overline{N}_{ij}$ are the numbers of reconstructed
$D^0$ and $\bar{D}^0$ decays, respectively, in bin $ij$. 
In order to avoid large statistical fluctuations near the phase space boundaries,
we calculate the asymmetry only for those bins having $N_{ij} + \overline{N}_{ij}>1000$.
This asymmetry is used to correct the tagged $K\pi$ events by
weighting each $D^0(\bar{D}^0)$ candidate falling into a valid bin with a weight
\begin{eqnarray}
  \nonumber
  && u_{D^0} = 1 - A^{\rm untag}_{\rm rec}(p_{D^0}, \cos\theta_{D^0})~,\\
  && u_{\bar{D}^0} = 1 + A^{\rm untag}_{\rm rec}(p_{\bar{D}^0}, \cos\theta_{\bar{D}^0})~.
\end{eqnarray}
Other candidates are discarded. 
The weighting applied to the tagged $K\pi$ decays results
in a measured $A_{\rm rec}^{\rm tag}$ free of all contributions in 
(\ref{eq3}) except for $A_{\epsilon}^\pi$. 

\begin{figure}
  \centerline{\epsfig{file=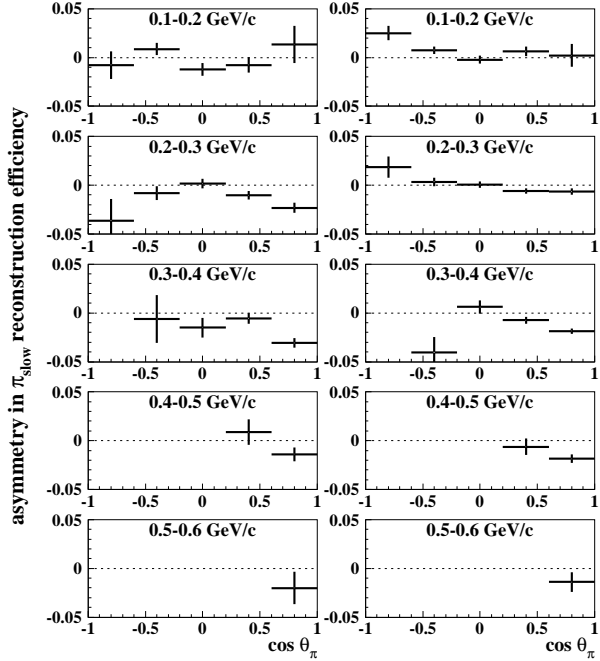,width=8cm}}
  \caption{Asymmetry of the slow pion efficiency, $A_{\epsilon}^\pi$, in momentum slices
    for the 3-layer (left) and the 4-layer (right) SVD configurations.}
  \label{pi_slow.eps}
\end{figure}

The slow pion asymmetry in bin $kl$ of the phase space
$(p_{\pi_s}, \cos \theta_{\pi_s})$ is thus determined with
\begin{equation}
  A_{\epsilon,kl}^\pi = \frac{n_{kl} - \overline{n}_{kl}}{n_{kl} + \overline{n}_{kl}}~,
\end{equation}
where $n_{kl}(\overline{n}_{kl})$ are the sums of weights of the $D^0(\bar{D}^0)$ 
candidates
falling in that bin. Again, we consider only bins with $n_{kl} + \overline{n}_{kl}>1000$.
The resulting asymmetry $A_{\epsilon}^\pi(p_{\pi_s}, \cos \theta_{\pi_s})$ determined 
in $5\times 5$ bins for the two 
SVD configurations is shown in Fig.~\ref{pi_slow.eps}. 
Averaging over the phase space the correction due to the slow pion
efficiency is found to be $(+0.76\pm 0.09)\%$.

The slow pion asymmetry is used to correct the $KK$ and $\pi\pi$ events.
The $D^0/\bar{D}^0$ candidates are weighted according to
\begin{eqnarray}
  \nonumber
  && w_{D^0} = 1 - A_\epsilon^\pi(p_{\pi_s}, \cos\theta_{\pi_s})~,\\
  && w_{\bar{D}^0} = 1 + A_\epsilon^\pi(p_{\pi_s}, \cos\theta_{\pi_s})~,
\end{eqnarray}
and only candidates in bins with valid  $A_\epsilon^\pi$ 
measurements are taken into account. This procedure results in a
corrected asymmetry $A_{\rm rec}^f$ of (\ref{eq2}), $A_{\rm rec}^{f,{\rm corr}}$, 
which is free of the
contribution due to the slow pion efficiency asymmetry. It is
calculated as 
\begin{equation}
  A_{\rm rec}^{f,{\rm corr}}(\cos \theta^*) = 
  \frac{m^f(\cos \theta^*) - \overline{m}^f(\cos \theta^*)}
       {m^f(\cos \theta^*) + \overline{m}^f(\cos \theta^*)}~,
\end{equation}
where $m^f(\overline{m}^f)$ represent the sum of weights of the $D^0 (\bar{D}^0)$
candidates in each bin of $\cos \theta^*$.
 
Finally, taking into account their specific dependence on $\cos{\theta^*}$, 
the asymmetries $A_{CP}$ and $A_{FB}$ are extracted by adding or subtracting 
bins at $\pm \cos{\theta^*}$:
\begin{eqnarray}
  \nonumber
  && A_{CP}^f=
  \frac{ A_{\rm rec}^{f,{\rm corr}}(\cos\theta^*)+A_{\rm rec}^{f,{\rm corr}}(-\cos\theta^*)}
       {2}~,\\
  && A_{FB}^f=
  \frac{ A_{\rm rec}^{f,{\rm corr}}(\cos\theta^*)-A_{\rm rec}^{f,{\rm corr}}(-\cos\theta^*)}
       {2}~.
\end{eqnarray}

\begin{figure}
  \centerline{\epsfig{file=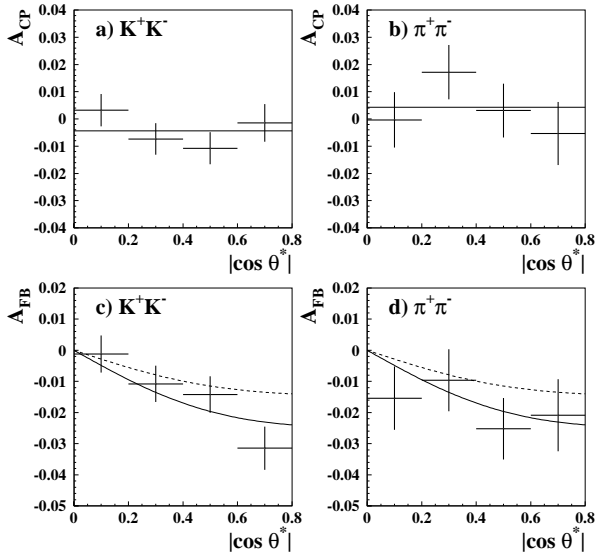,width=8cm}}
  \caption{$CP$-violating asymmetries in (a) $KK$ and (b) $\pi\pi$ final states,
    and forward-backward asymmetries in (c) $KK$ and (d) $\pi\pi$ final states.
    The solid curves represent the central values obtained from the least square 
    minimizations; the dashed curves in (c) and (d) show the leading order expectation.}
  \label{final_plot.eps}
\end{figure}

The results are presented in Fig.~\ref{final_plot.eps}. By fitting a constant to 
the $A_{CP}^f$ data points we obtain results consistent with no $CP$ violation:
\begin{eqnarray}
  \nonumber
  A_{CP}^{KK} &=& (-0.43 \pm 0.30)\%~, \\
  A_{CP}^{\pi\pi} &=& (+0.43 \pm 0.52)\%~.
\end{eqnarray}
The errors are statistical only; however, the statistical uncertainties of the slow
pion corrections are not included. The forward-backward asymmetry $A_{FB}$
decreases with $\cos \theta^*$ and has a value $\approx -3\%$ at $\cos \theta^*=0.8$;
results from the two samples are consistent. At leading order, 
the asymmetry at this energy is expected to be
$A_{FB}^{c\bar{c}}(\cos \theta^*) = a^{c\bar{c}} \cos \theta^*/(1+\cos^2 \theta^*)$,
with $a^{c\bar{c}}=-2.9\%$~\cite{PDG2004}. A simultaneous fit to the
two samples yields an acceptable goodness-of-fit ($\chi^2/n_{\rm dof} = 4.5/7$)
and $a^{c\bar{c}}=(-4.9 \pm 0.8)\%$, where the error
is statistical (see Fig.~\ref{final_plot.eps}). 

\section{Systematics}

The experimental procedure was checked using the generic continuum MC simulation;
the resulting $A_{CP}$ and $A_{FB}$ were found to be in a good agreement 
with the generated values. We also tested for possible bias in the result 
by re-weighting MC samples with several non-zero $A_{CP}$ values; no significant
bias was found.  
 
We consider three sources of systematic uncertainty to be significant 
(Table~\ref{syst-sum.tab}). The first source is the mass-sideband subtraction 
procedure used for signal counting.
Possible systematic uncertainties arise due to the 
difference in signal shapes of $D^0$ and $\bar{D}^0$ candidates and due to the
possible difference in the background between the signal window and sideband. The former
source can introduce an additional asymmetry if the signal window is not sufficiently
wide. We observe small but significant differences in the $q$ signal shape 
of the tagged samples.  
By studying the normalized (in order to asses only the effect of the shape
difference) $q$ distributions of the tagged
$D^0(\bar{D}^0)\to K\pi$ samples we estimate the systematic uncertainty of this source 
to be 0.02\% (0.04\%) for the $KK$ ($\pi\pi$) sample.
To account for a possible difference in backgrounds we vary
the position of the sideband. We find 0.01\% ($KK$) and 0.03\% ($\pi\pi$) variations
in the result. Background due to a correctly reconstructed $D^0$ candidate combined
with a random slow pion is not removed by the $M$ sideband subtraction.
Its fraction (0.6\%) is
estimated from the tuned MC simulation. The possible asymmetry induced
by this type of background is estimated from the $q$
sideband to be at most 0.03\%.

The second source of systematic error is the slow pion efficiency correction. The 
statistical errors on $A_{\epsilon}^\pi(p_{\pi_s}, \cos \theta_{\pi_s})$ 
contribute an uncertainty of 0.09\%.
The impact of binning of the slow pion asymmetry is studied by producing maps 
with three different choices of bin sizes
($10\times 10,~20\times 20,~50\times 50$ for 
$A^{\rm untag}_{\rm rec}$, and
$5\times 5,~10\times 10,~20\times 20$ for $A_\epsilon^\pi$) and repeating
the procedure for extracting $A_{CP}$. We find 0.03\% ($KK$) and 0.02\% ($\pi\pi$)
variations in the result. The minimum required number of events per bin is varied from
100 to 10000, and the resulting variation
in $A_{CP}$ is 0.04\% (0.03\%) for the $KK$ ($\pi\pi$) sample.

The third source of systematic uncertainty is the $A_{CP}$ extraction procedure. 
By varying
the binning in $|\cos \theta^*|$ we obtain a 0.03\% variation in the result. We change
the treatment of the running periods with 3- and 4-layer SVD configuration;
we find an 0.01\% (0.02\%) change in the result for the $KK$ ($\pi\pi$) sample.

Finally, we add the individual contributions in quadrature
to obtain the total systematic uncertainty. The result is
0.11\% (0.12\%) for the $KK$ ($\pi\pi$) sample. The dominant
source is the statistical uncertainty on $A_{\epsilon}^\pi$, and
thus the majority of the systematic error will decrease when
a larger $K\pi$ data sample is available.

\begin{table}
  \caption{Summary of systematic uncertainties in $A_{CP}$.}
  \label{syst-sum.tab}
  \begin{center}
    \begin{tabular}{lcc}
      \hline
      Source & $D^0 \to K^+K^-$~~ & $D^0 \to \pi^+\pi^-$ \\
      \hline
      Signal counting        & 0.04\% & 0.06\% \\
      Slow pion corrections  & 0.10\% & 0.10\% \\
      $A_{CP}$ extraction     & 0.03\% & 0.04\% \\
      \hline
      Sum in quadrature       & 0.11\% & 0.12\% \\
      \hline
    \end{tabular}
  \end{center}
\end{table}

\section{Conclusions}

We measure time-integrated $CP$-violating asymmetries $A_{CP}$ 
in decays to $CP$ eigenstates
$D^0 \to K^+K^-$ and $D^0 \to \pi^+\pi^-$ using 540~fb$^{-1}$ of data.
The detector-induced asymmetries are corrected with a precision of 0.1\%
by using tagged and untagged $D^0 \to K^-\pi^+$ decays. We obtain:
\begin{eqnarray}
  \nonumber
  && A_{CP}^{KK} = (-0.43 \pm 0.30 \pm 0.11)\%~, \\
  \nonumber
  && A_{CP}^{\pi\pi} = (+0.43 \pm 0.52 \pm 0.12)\%~, \\
  && A_{CP}^{KK}-A_{CP}^{\pi\pi}=(-0.86\pm 0.60\pm 0.07)\%.
  \label{acp-final}
\end{eqnarray}
The results show no evidence for $CP$ violation and agree
with SM predictions. In (\ref{acp-final}) we also list the difference
$A_{CP}^{KK}-A_{CP}^{\pi\pi}$, which is calculated by treating the systematic
errors arising from the slow pion corrections and $A_{CP}$ extraction
as fully correlated between the two modes. A significant difference
between the measured asymmetries in the $KK$ and $\pi\pi$ modes would
be a sign of direct CPV (Eq.~(\ref{eq1})).

To determine the direct CPV asymmetries $a_d^f$ of (\ref{eq1}),
the results in (\ref{acp-final}) can be compared to the result for
the indirect CPV asymmetry in Ref.~\cite{belle_mix}. While the selected data
samples of $D^0\to K^+K^-, \pi^+\pi^-$ in the two measurements are
almost identical, the methods of extracting the $CP$ violating
asymmetries depend on different observables and hence the statistical
uncertainties are uncorrelated. The same holds also for the systematic
errors. The direct CPV asymmetries following from the sum of
$A_{CP}^f$ and $A_\Gamma$ are
\begin{eqnarray}
  \nonumber
  a_d^{KK} &=& (-0.42\pm 0.42\pm 0.19)\%~,\\
  a_d^{\pi\pi}~ &=& (+0.44\pm 0.60\pm 0.19)\%~.
\end{eqnarray}
The measurement uncertainties are above the level of the expected asymmetry 
in the SM. 

We also measure the forward-backward asymmetry in the production of
$D^{*+}$ that arises from the underlying asymmetry in the $e^+e^-\to
c\bar{c}$ process. The asymmetry agrees with the form 
$A_{FB}^{c\bar{c}}(\cos \theta^*) = a^{c\bar{c}} \cos \theta^*/(1+\cos^2 \theta^*)$
expected at leading order, but we find $a^{c\bar{c}}=(-4.9 \pm 0.8)\%$,
larger than the leading-order value of $-2.9\%$. Radiative and other (hadronic) 
corrections are expected to cause the effective $a^{c\bar{c}}$ to deviate 
from its leading-order value.

\section*{Acknowledgments}

We thank the KEKB group for excellent operation of the
accelerator, the KEK cryogenics group for efficient solenoid
operations, and the KEK computer group and
the NII for valuable computing and SINET3 network
support.  We acknowledge support from MEXT and JSPS (Japan);
ARC and DEST (Australia); NSFC (China); 
DST (India); MOEHRD, KOSEF and KRF (Korea); 
KBN (Poland); MES and RFAAE (Russia); ARRS (Slovenia); SNSF (Switzerland); 
NSC and MOE (Taiwan); and DOE (USA).

\end{document}